\documentstyle[12pt]{article}
\date{}
\begin{document}
\title{{\LARGE\sf Blocking and Persistence in the Zero-Temperature Dynamics
of Homogeneous and Disordered Ising Models}}
\author{
{\bf C. M. Newman}\thanks{Partially supported by the 
National Science Foundation under grant DMS-98-02310.}\\
{\small \tt newman\,@\,cims.nyu.edu}\\
{\small \sl Courant Institute of Mathematical Sciences}\\
{\small \sl New York University}\\
{\small \sl New York, NY 10012, USA}
\and
{\bf D. L. Stein}\thanks{Partially supported by the 
National Science Foundation under grant DMS-98-02153.}\\
{\small \tt dls\,@\,physics.arizona.edu}\\
{\small \sl Depts.\ of Physics and Mathematics}\\
{\small \sl University of Arizona}\\
{\small \sl Tucson, AZ 85721, USA}
}
\maketitle
\begin{abstract}
A ``persistence'' exponent $\theta$ has been extensively used to describe
the nonequilibrium dynamics of spin systems following a deep quench: for
zero-temperature homogeneous Ising models on the $d$-dimensional
cubic lattice $Z^d$, the fraction $p(t)$ of
spins not flipped by time $t$ decays to zero like $t^{-\theta(d)}$ for low
$d$; for high $d$, $p(t)$ may decay to $p(\infty)>0$, because of ``blocking''
(but perhaps still like a power).  What are the effects of disorder or
changes of lattice? We show that these can quite generally lead to blocking
(and convergence to a metastable configuration) even for low $d$, and then
present two examples --- one disordered and one homogeneous --- where
$p(t)$ decays {\it exponentially\/} to $p(\infty)$.
\end{abstract}
\small
\normalsize

In modelling the nonequilibrium dynamics of spin systems following a deep
quench, the following question naturally arises
\cite{Stauffer,Derrida95,DHP,MH,DOS,MS,NT,MB}:
given a spin system at zero temperature with random starting configuration
and evolving according to the usual Glauber dynamics, what is the
probability $p(t)$ at time $t$ that a spin has not yet flipped?

For the homogeneous ferromagnetic Ising model on $Z^d$,
this probability has been found to decay at large times as
a power law $p(t)\sim t^{-\theta(d)}$ \cite{Stauffer,Derrida95,DHP} for
$d<4$.  The ``persistence'' exponent $\theta(d)$ is considered to be a new
universal exponent governing nonequilibrium dynamics following a deep
quench \cite{MS}.  The persistence problem can be extended to positive
temperatures by considering the dynamics of the local order parameter
rather than that of single spins \cite{MBCS,CS,DG}.

In this paper we confine our attention to dynamics at zero temperature in
infinite spin systems.  In the usual case of asynchronous updating, a spin
is chosen at random (this can be made precise for infinite systems, 
as in \cite{NS99}) 
and then: always flips if the resulting
configuration has lower energy, never flips if the energy is raised, and
flips with probability $1/2$ if the resulting energy change is zero.  We
will consider these dynamics for random initial configurations $\sigma^0$
(in which each spin is equally likely to be up or down, independent of the
others) in both disordered ferromagnets and spin glasses with continuous
coupling distributions, and also for uniform ferromagnets on lattices other
than $Z^d$ (e.g., hexagonal lattices in $2D$).

Our first result is that the persistence phenomenon as described
above is unstable to the introduction of randomness into the
spin couplings, or even to some changes in lattice structure.  
For the random ferromagnet, spin glass, $2D$ hexagonal
ferromagnet, and others to be discussed below, we will see that a positive
fraction of spins never flip and every spin flips only finitely many 
times.

The ``frozenness'' of a nonvanishing fraction of spins (sometimes
referred to as ``blocked'' spins \cite{DOS}) has been reported in numerical
simulations of Ising ferromagnets on $Z^d$ with $d>4$ \cite{Stauffer} and
$q$-state Potts models on square lattices for $q>4$ \cite{DOS}.  The
problem can then be recast by restricting attention to only those spins
that eventually do flip, and asking for the conditional probability that
such spins haven't yet flipped by time $t$.  Simulations of Potts
models appeared to indicate that this probability (proportional to
$p(t)-p(\infty)$) also decays as a power law at long times \cite{DOS}
(however, some curvature on their log-log plots was noted).

We will examine the same question for disordered Ising systems and also for
homogeneous systems that show blocking.  Although we cannot yet answer
this question in general, we will present calculations on two systems, the
homogeneous ferromagnet on a quasi-$1D$ ``ladder'' and the
$1D$ disordered spin chain \cite{FDM}, showing that
$p(t)-p(\infty)$ decays {\it exponentially\/} as $t\to\infty$.
Exponential decay for $d\ge 2$ will also be discussed.

\medskip

{\it Persistence and local nonequilibration.}  The analysis of persistence
exponents suggests that the fraction of sites that remain in the same phase
(for $T>0$) or spin value (at $T=0$) from time $t_1$ to time $t_2$ tends to
zero for $1\ll t_1\ll t_2$.  It therefore implies the presence of
{\it local nonequilibration\/} (LNE) \cite{NS99}: that in any fixed, finite
region, there exists no finite time after which the spins within remain in
a single phase; that is, domain walls forever sweep across the region.  At
zero temperature, the presence of LNE means every spin flips infinitely
often (in almost every sample).

Why does the decay to zero of $p(t)$ (coming from the analysis of
persistence exponents at $T=0$) suggest that every spin flips infinitely
many times?  Suppose instead that a positive fraction of spins flip only
finitely many times.  Then it is reasonable to expect that a (smaller but
still positive) fraction of spins {\it never\/} flip, and $p(t)$ would not
decay to zero.  While not proved in general, this argument applies to all
systems treated here.

It was proved in \cite{NNS} (see also \cite{NS99}) that, in the homogeneous
Ising ferromagnet on the square lattice with a random initial spin
configuration, every spin indeed flips infinitely often at zero
temperature, consistent with persistence results in the literature.
(Similar results apply to several other systems, and can be extended to
positive temperature with the local order parameter in a region replacing
individual spins \cite{NS99}.)

\medskip

{\it Blocking.}  What about the zero-temperature dynamics of systems with
continuous disorder?  In any dimension and on any lattice, it can be proved
for these (and many other) systems that {\it every spin flips only finitely
many times.\/} These systems exhibit ``blocking'' and for them
$p(t)$ does not decay to zero.

These are examples of a general result \cite{NNS} applying to the dynamical
evolution (following a deep quench) of infinite-volume Ising spin systems
with Hamiltonian
\begin{equation}
\label{eq:Hamiltonian}
{\cal H}=-\sum_{<xy>} J_{xy}\sigma_x\sigma_y\ ,
\end{equation}
where the sum is over nearest neighbors.  If the distribution of couplings
is continuous with finite mean, then it can be proved that every spin flips
only finitely many times (for almost every $\sigma^0$, realization $\omega$
of the dynamics and realization ${\cal J}$ of the couplings).

The proof of this theorem yields a more general result that shows that,
even without the continuity assumption on the distribution of couplings,
for almost every ${\cal J}$, $\sigma^0$, and $\omega$, there can be only
finitely many flips of any spin that cause a {\it nonzero\/} energy change.
This is why the above result applies to ordinary spin glasses and random
ferromagnets with a continuous distribution of couplings (e.g., Gaussian or
uniform): the probability of a ``tie'' in any sum or difference of a given
spin's nearest-neighbor coupling strengths (and therefore the probability
of a spin flip costing zero energy) is zero, and the result follows.

We sketch the proof here; for further details, see
\cite{NNS}.  Let $\sigma_x^t$ be the value of $\sigma_x$ at time $t$ 
for fixed $\omega$, $\sigma^0$ and ${\cal J}$.  Let  
\begin{equation}
\label{eq:energy}
E(t)=-(1/2)\overline{\sum_{y:|x-y|=1}J_{xy}\sigma_x^t\sigma_y^t}
\end{equation}
where the bar indicates an average over ${\cal J}$, $\sigma^0$, and
$\omega$.  By translation-ergodicity of the distributions from which ${\cal
J}$, $\sigma^0$, and $\omega$ were chosen, and using the assumption that
$\overline{|J_{xy}|} < \infty$,
it follows that $E(t)$
exists, is independent of $x$, and equals the energy density (i.e., the
average energy per site) at time $t$ in almost every realization of ${\cal
J}$, $\sigma^0$, and $\omega$.

Because every spin flip lowers the energy, $E(t)$ monotonically decreases
in time (note that $E(0)=0$) and has a finite limit $E(\infty)$ ($\ge
-d\overline{|J_{xy}|}$).  Now choose any fixed number $\epsilon>0$, and let
$N_x^\epsilon$ be the number of spin flips (over all time) of the spin at
$x$ that lower the energy by an amount $\epsilon$ or greater.  Then
$-\infty<E(\infty)\le -\epsilon\overline{N_x^\epsilon}$ so that for every
$x$ and $\epsilon > 0$, $N_x^\epsilon$ is finite.  Let $\epsilon_x$ be the
minimum energy (magnitude) change resulting from a flip of $\sigma_x$; then
although $\epsilon_x$ varies (differently in each ${\cal J}$) with $x$, it
is sufficient that it is strictly positive.

This result applies also to homogeneous systems on certain lattices, such
as Ising ferromagnets on lattices with an {\it odd\/} number of
nearest-neighbors so that ties in energy cannot occur.  Such lattices
include the hexagonal (or honeycomb) lattice in $2D$, and 
the double-layered cubic lattices
$Z^d\times \{0,1\}$ (i.e., a ``ladder'' when $d=1$, two horizontal planes
separated by unit vertical distance when $d=2$, and so on) \cite{note2}.

As for blocking in these systems, it is 
elementary to show that a positive fraction of
spins will never flip.  Consider first the hexagonal lattice.  If the spins
on any single hexagon are all up or all down, they will form a stable
configuration that will never change.  Such configurations (and of course
similar larger-scale ones) occur with positive density in almost every
$\sigma^0$.  Similarly, in the ladder, any square
with all spins up or down is stable.  The extension to general $Z^d\times
\{0,1\}$ is straightforward.

Turning to disordered systems, consider first the random ferromagnet on 
$Z^2$.  For almost every ${\cal J}$, there will be a positive
density of plaquettes whose couplings satisfy the following: on
each of the four corners, the sum of the two couplings that connect to
adjacent corners of the square is greater than the sum of the two couplings
to sites outside the square.  If the spins at the four corners initially
are all up or all down, the spin configuration on the square will again be
stable.  A similar construction can be used for general $d$ and for spin
glasses.

To summarize, our first result has been to prove that many ordered and
disordered spin systems display two important zero-temperature dynamical
properties, which when taken together lead them to exhibit novel
persistence behavior.  The first concerns the presence of blocking, meaning
a positive fraction of spins never flip.  In the systems we treat, this is
a zero time property in that some of the spins are blocked by the nature of
$\sigma^0$ (and ${\cal J}$), regardless of the dynamics realization. The
second property concerns infinite time: the existence of a limiting
(metastable) spin configuration $\sigma^\infty$, since every spin flips
only finitely many times.  Although the second of these properties probably
implies the first, the first does not imply the second \cite{GNS}.  Our
next result is to show that for at least some of these systems, these two
properties lead to an exponential (as opposed to power law) decay of the
quantity $p(t)-p(\infty)$ at large times.

\medskip

{\it Exponential decay.\/} In this section we study the large-time
behavior of $p(t)-p(\infty)$, the probability that a spin will flip at some
time but has not yet flipped by time $t$.  We will prove that this quantity
decays exponentially by showing the same for the larger probability
$\tilde{p}(t)$ that a spin will flip at some time after $t$ (whether or not
it has flipped before).  We consider two systems, one homogeneous
(the uniform Ising ferromagnet on the ladder) and one
disordered (the $1D$ continuously disordered Ising chain).

Consider first a homogeneous system where every site has an odd number
$M$ of neighbors. (Systems such as $\pm J$ spin glasses where the
{\it signs\/} are disordered but not the $|J_{xy}|$'s also fall into this
category of examples.) Consider at time $\tau$, all sites $y$ such that
the spin at $y$ will flip after time $\tau$, and denote by 
${\cal C}_x (\tau)$ the cluster of such sites that contains $x$
(an empty cluster if the spin at $x$ will not flip after time $\tau$).
We will show below for the ladder model that with $\tau = 0$, the
distribution of the number of sites $|{\cal C}_x (\tau)|$ in these clusters
has an exponential tail; i.e., the probabilities for large cluster
sizes are bounded by:
\begin{equation}
\label{eq:size}
P(|{\cal C}_x (\tau)| \ge n )\, \le \, A\,e^{-kn}
\end{equation}
for some $A < \infty$ and $k > 0$. We next show that this implies exponential
decay of $\tilde{p}(t)$.

Since each flip in ${\cal C}_x (\tau)$ lowers the energy of that cluster by
at least $2$ and since the total energy of the cluster 
lies somewhere between
$-M |{\cal C}_x (\tau)|$ and $M |{\cal C}_x (\tau)|$ 
(we take $J=1$ here), it follows that the entire
cluster must reach its final configuration after no more than $M |{\cal
C}_x (\tau)|$ flips. Let $T_1$ denote the (random) amount of time after
$\tau$ until the first flip in ${\cal C}_x (\tau)$, $T_2$ the amount of
time after $\tau + T_1$ until the second flip, etc.  Clearly, as long as
flips are possible, the $T_i$'s are bounded above by independent
exponential (mean one) random variables $T_i'$. Thus the time of the last
flip of $x$ is bounded above by $\tau + T_1' + \cdots + T_{M|{\cal C}_x
(\tau)|}'$ and so for $t > \tau$,
\begin{equation}
\label{eq:bound}
p(t) - p(\infty)  \le  \tilde{p}(t)  
\le  \sum_{n=1}^\infty P(|{\cal C}_x (\tau)|=n)
P(T_1' + \cdots + T_{Mn}' \ge t-\tau)\, . 
\end{equation}
The probability density of $T_1' + \cdots + T_j'$ is $f(s) =
s^{j-1}e^{-s}/(j-1)!$ and so for $t > \tau$,
\begin{eqnarray}
\label{eq:bound2}
p(t) - p(\infty) & \le & \sum_{n=1}^\infty A e^{-kn} \int_{t-\tau}^\infty
[s^{Mn-1} / (Mn-1)!]e^{-s}ds \nonumber\\
&\le& \int_{t-\tau}^\infty 
\sum_{j=1}^\infty A(e^{-k/M})^j [s^{j-1}/(j-1)!]e^{-s}ds \nonumber\\
& = & Ae^{-k/M}\int_{t-\tau}^\infty \exp(e^{-k/M} s - s) ds \nonumber\\
&=& A' e^{-k't}
\end{eqnarray}
where the constants $A'$ and $ k'$ depend on $A, k, M$ and $\tau$.

It remains to show that (\ref{eq:size}) is valid in the ladder ferromagnet
with $\tau = 0$. (Similar arguments work for the ladder antiferromagnet or
$\pm J$ spin glass.) To do this we note that a single plaquette has an
initial probability $p_0=1/8$ of having its four corner spins all plus or
all minus (we call such a blocked
plaquette ``frigid'').  The lattice's sites take
integer values $(x_i,y_i)$, with $-\infty<x_i<\infty$ and $y_i=0,1$.  A
lower bound for the initial number of frigid plaquettes can be obtained by
considering only those plaquettes whose left edges occur at even $x_i$ (we
define the location of a plaquette by the position of its left edge); such
plaquettes do not overlap and so their probabilities of being frigid are
independent.  If the plaquette at the origin is frigid, and $P(0\rightarrow
2n)$ is the probability that there is no frigid plaquette between 0 and
$x_{2n}$, then
\begin{equation}
\label{eq:frozen}
P(0\rightarrow 2n)\le(1-p_0)^{n}=\exp(-kn)\ ,
\end{equation}
where $k=|\log(1-p_0)|$.  So the ladder is broken up into finite segments,
bounded to either side by a frigid plaquette, whose length distribution has
an exponential tail.  This yields Eq.~(\ref{eq:size}).

Our second example is a disordered $1D$ spin chain in zero field.  The
analysis is essentially the same for either the spin glass or the random
ferromagnet, so for specificity we study a ferromagnet whose couplings $J_z
\equiv J_{z,z+1}$ are independent random variables
taken from the uniform distribution on $[0,1]$.  The key idea here is that
the chain breaks up into finite, disjoint ``influence segments'' whose
union is the infinite chain.  An influence segment is a dynamical construct
defined (for a given ${\cal J}$) as follows: two
sites $x$ and $y$ belong to the same influence segment if and only if
either the state of $\sigma_x$ can dynamically induce a change in the state
of $\sigma_y$ or vice-versa (or both).  To illustrate,
suppose that the coupling $J_x$ is larger than both $J_{x-1}$ and
$J_{x+1}$; i.e., $J_x$ is a local maximum.  Then it is clear that the state
of $\sigma_x$ can be dynamically influenced by $\sigma_{x+1}$ but not by
$\sigma_{x-1}$ (and similarly, the state of $\sigma_{x+1}$ can be
dynamically influenced by $\sigma_x$ but not by $\sigma_{x+2}$).  That is,
no state of the spin $\sigma_{x-1}$ can alter the sign of the energy change
$\Delta{\cal H}_x$ that would result from a flip of $\sigma_x$.  To
summarize, two sites $x$ and $y$ are defined to be in the same influence
cluster if and only if either $\sigma_x$ can influence $\sigma_y$, or
$\sigma_y$ can influence $\sigma_x$, or both \cite{NN}.

Influence segments for the disordered $1D$ chain are then constructed as
follows \cite{NNS}.  Consider the doubly infinite sequences $x=x_m$ of
sites where $J_x$ is a local maximum and $y=y_m\in(x_m,x_{m+1})$ where
$J_y$ is a local minimum:  the couplings are strictly increasing from
$y_{m-1}$ to $x_m$ and strictly decreasing from $x_m$ to $y_m$.  The set of
spins at the sites $\{y_{m-1}+1, y_{m-1}+2,\ldots y_m\}$ determines a
single influence segment.

To see this, note that the spin at $y_m$ cannot influence the one at
$y_m+1$ (or vice-versa); similarly, at the other end $y_{m-1}+1$ cannot
influence $y_{m-1}$.  Now consider the spins at $x_m$ and $x_m+1$, which
are within the interval $\{y_{m-1}+1, y_{m-1}+2,\ldots y_m\}$.  Clearly,
the spin at $x_m-1$ can never influence the spin at $x_m$, and the spin at
$x_m+2$ can never influence the one at $x_m+1$.  So once the spins at $x_m$
and $x_m+1$ agree (either initially in $\sigma^0$ or after either spin
flips) the final value of every spin in the interval $\{y_{m-1}+1,
y_{m-1}+2,\ldots y_m\}$ is determined through a ``cascade'' of influence to
either side of $\{x_m,x_m+1\}$ (which is put into effect as the Poisson
clocks successively ring) until $y_{m-1}+1$ and $y_m$, respectively, are
reached.

Given this, the analysis leading to Eqs.~(\ref{eq:bound}) and
(\ref{eq:bound2}) applies as before.  One needs only an estimate analogous
to Eq.~(\ref{eq:size}) for the probability distribution of influence
segment sizes.  In fact, the decay here for large size $n$ is faster than
in Eq.~(\ref{eq:size}); the probability of $n$ independent coupling random
variables being ordered so as to have a single local maximum falls off as
$1/n!$ (times an exponential factor).

In these two examples different factors determine the distribution of
dynamical cluster sizes: for the homogeneous ferromagnet on the ladder,
they're determined by the initial spin configuration $\sigma^0$; for the
disordered $1D$ chain, they're determined by the coupling realization
${\cal J}$.

In this section we considered two examples, one ordered and one disordered,
but both one-dimensional (or quasi-one-dimensional).  There is another
system that shows the same behavior in {\it any\/} dimension: the highly
disordered spin glass (or ferromagnet) \cite{NS94,BCM,NS96}.  Using similar
arguments, this system can also be shown to display an exponential decay to
its final state \cite{Nthesis}.  We expect that a related model, in which
coupling magnitudes are ``stretched'' in the manner of references
\cite{NS94,BCM,NS96} but only up to a finite length scale, would show
similar behavior.  This last model is of interest because its thermodynamic
behavior is expected to be similar to that of the ordinary spin glass (or
random ferromagnet).

\medskip

{\it Discussion.\/} Most work on persistence at zero temperature has
examined systems, such as the homogeneous Ising ferromagnet on $Z^d$ in low
$d$, where the quantity $p(t)$ decays to zero as a power law.  We
have shown here that there is a second class of models in which
$p(\infty)>0$: these include systems with continuous disorder and
homogeneous systems on other lattices.  In several of these the persistence
decay is exponential rather than power law.  It would be of interest to see
whether this fast decay holds in other systems in this general class, such
as the $2D$ homogeneous ferromagnet on a hexagonal lattice 
\cite{DH} or an ordinary
spin glass with $d > 1$.

Although we don't know the answer in the general case, we can speculate
using a rough argument that the answer may be yes.  If every spin flips
only finitely often, as time progresses an increasing number of spins will
``freeze''; i.e., they cease to flip.  It is reasonable to expect that
after some finite time ``unfrozen'' spins no longer percolate, so that the
dynamics is confined to noninteracting finite clusters, as in the examples
treated here.  Of course, there remain serious gaps: this is not
independent percolation, and the dynamics in the localized clusters (should
they exist) would need to be worked out, so the conclusion should
be treated with caution. 

There is a third class of systems not discussed here; in these, a
positive fraction of spins flip infinitely often and a positive fraction
flip only finitely many times.  One such system is the two-dimensional $\pm
J$ spin glass \cite{GNS}.  Although it appears that
$p(\infty)>0$, determining the large-time behavior of $p(t)-p(\infty)$
remains an open problem.

{\it Acknowledgments.\/} This research was partially supported by NSF
Grants DMS-98-02310 (CMN) and DMS-98-02153 (DLS).

\renewcommand{\baselinestretch}{1.0}


\small
\begin{thebibliography}{10}

\bibitem{Stauffer}
D.~Stauffer, J.~Phys.~A {\bf 27}, 5029 (1994).

\bibitem{Derrida95}
B.~Derrida, J.~Phys.~A {\bf 28}, 1481 (1995).

\bibitem{DHP}
B.~Derrida, V.~Hakim, and V.~Pasquier, Phys.~Rev.~Lett. {\bf 75}, 751
(1995).

\bibitem{MH}
S.N.~Majumdar and D.A.~Huse, Phys.~Rev.~E {\bf 52}, 270 (1995).

\bibitem{DOS}
B.~Derrida, P.M.C.~de~Oliveira, and D.~Stauffer, Physica {\bf 224A},
604 (1996).

\bibitem{MS}
S.N.~Majumdar and C.~Sire, Phys.~Rev.~Lett.~{\bf 77}, 1420 (1996).
Note that in this paper the definition of $\theta$ differs by a factor
of 2 from that used here.

\bibitem{NT}
T.J.~Newman and Z.~Toroczkai, Phys.~Rev.~E {\bf 58}, 2685 (1998).

\bibitem{MB}
S.N.~Majumdar and A.J.~Bray, Phys.~Rev.~Lett.~{\bf 81}, 2626 (1998).


\bibitem{MBCS}
S.N.~Majumdar, A.J.~Bray, S.J.~Cornell, and C.~Sire,  Phys.~Rev.~Lett.~{\bf
77}, 3704 (1996).

\bibitem{CS}
S.~Cuelle and C.~Sire, J.~Phys.~A {\bf 30}, L791 (1997).

\bibitem{DG}
J.-M.~Drouffe and C.~Godr\'eche, ``Stationary definition of persistence
for finite temperature phase ordering'', preprint (1998).


\bibitem{NS99}
C.M.~Newman and D.L.~Stein, J.~Stat.~Phys.~{\bf 94}, 709 (1999).

\bibitem{FDM} Results on persistence in some random $1D$ systems with $T>0$
appear in D.S.~Fisher, P.~Le~Doussal, and C.~Monthus, Phys.~Rev.~Lett.~{\bf
80}, 3539 (1998).

\bibitem{NNS}
S.~Nanda, C.M.~Newman, and D.L.~Stein, ``Dynamics of Ising spin systems at
zero temperature'', in {\it On Dobrushin's Way 
(from Probability Theory to Statistical Physics)\/}, R.~Minlos, 
S.~Shlosman and Y.~Suhov, eds.~(Am.~Math.~Soc., Providence) (to be published).

\bibitem{note2} One would get the same result for any model with a change
of dynamical rule that eliminates ties altogether, as in certain voter
models.  However, dynamical rules of this kind are not natural $T\to 0$
limits of the usual positive-temperature Glauber dynamics.

\bibitem{GNS}
A.~Gandolfi, C.M.~Newman, and D.L.~Stein, ``Zero-Temperature Dynamics
of $\pm J$ Spin Glasses and Related Models'', in preparation.

\bibitem{NN} A general discussion of influence percolation is
given in S.~Nanda and C.M.~Newman, ``Random Nearest Neighbor and Influence
Graphs on $Z^d$'', in Random Structures and Algorithms
(to be published); see also Ref.~\cite{NNS}.

\bibitem{NS94}  C.M.~Newman and D.L.~Stein, 
Phys.~Rev.~Lett.~{\bf 72}, 2286 (1994).

\bibitem{BCM}  J.R.~Banavar, M.~Cieplak and A.~Maritan, 
Phys. Rev. Lett.~{\bf 72}, 2320 (1994).

\bibitem{NS96}  C.M.~Newman and D.L.~Stein, 
J.~Stat.~Phys.~{\bf 82}, 1113 (1996).

\bibitem{Nthesis}  S.~Nanda, ``Spatial Random Graphs and Dynamics
of Disordered Systems'', Ph.D.~Dissertation, New York Univ., May, 1998.

\bibitem{DH} Evidence for fast decay in this and related systems is
mentioned in C.D.~Howard, ``Zero-temperature Ising spin dynamics on the
homogeneous tree of degree three'', preprint (1999).

\end{thebibliography}
\end{document}